# Enhancement of Carrier Mobility in Semiconductor Nanostructures by Carrier Distribution Engineering

Binxi Liang (梁斌熙), Luhao Liu (刘陆豪), Jiachen Tang (唐家晨), Jian Chen (陈健), Yi Shi (施毅)*, Songlin Li (黎松林)*

*School of Electronic Science and Engineering, National Laboratory of Solid-State Microstructures and Collaborative Innovation Center of Advanced Microstructures, Nanjing University, Nanjing, Jiangsu 210023, China*

*Email: sli@nju.edu.cn and yshi@nju.edu.cn

**Abstract:** Two-dimensional (2D) van der Waals semiconductors are appealing for low-power transistors. Here, we show the feasibility in enhancing carrier mobility in 2D semiconductors through engineering the vertical distribution of carriers confined inside the ultrathin channels via symmetrizing gate configuration or increasing channel thickness. Through self-consistently solving the Schrödinger-Poisson equations, the shapes of electron envelope functions are extensively investigated by clarifying their relationship with gate configuration, channel thickness, dielectric permittivity, and electron density. The impacts of electron distribution variation on various carrier scattering matrix elements and overall carrier mobility are insightfully clarified. It is found that the carrier mobility can be generally enhanced in the dual-gated configuration due to the centralization of carrier redistribution in the nanometer-thick semiconductor channels and the rate of increase reaches up to 23% in $HfO_2$ dual-gated 10-layer $MoS_2$ channels. This finding represents a viable strategy for performance optimization in transistors consisting of 2D semiconductors.

## Introduction

The isolation of graphene and relevant two-dimensional (2D) van der Waals semiconductors has stimulated the interest and exploration of their applications as low-power field-effect transistors (FETs) for more-Moore electronics[1–4] and as driving elements in displays.[5–7] In addition to the merit of atomic thickness that is favorable for enhancing gate electrostatic control,[8–11] the self-saturated surfacial atomic structure represents another important advantage superior to silicon for attaining high device performance by suppressing extrinsic charge scattering events.[11–14] Even though 2D semiconductor FETs have been widely investigated,[15–19] the fine physical studies on the impacts of device parameters and gate configuration on the fundamental envelope function, $\Psi(z)$, or distribution of electrons in the ultrathin channels of 2D FETs and on the carrier mobility remain insufficient.

For 2D FETs, there are various surface-associated carrier scattering mechanisms sensitive to the vertical electron distribution inside ultrathin channels.[20–23] Thus, an accurate description of $\Psi(z)$ and its variation with external modulations is essential to reliably calculate the electron scattering rates and overall device performance. For electrically gated bulk Si FETs, the energy bands of the conduction channels bend down at the Si-SiO$_2$ interface, where the electrically induced 2D electrons are confined within nanometers ranges,[24] forming a ~10 nm thick inversion layer. The empirical $\Psi(z)$ of 2D electrons in the FET inversion layer was first proposed by Fang and Howard,[25] which works well for bulk semiconductors but invalid for the ultrathin 2D van der Waals materials. In this context, Kim et al fell back on basic physical principles to calculate $\Psi(z)$ in the 2D semiconductors by self-consistent Schrödinger-Poisson solution[26] and, in the samples that are thicker than 20 nm, they derived a distribution similar to bulk silicon. For semiconductors even thinner or atomically thin, most researchers, for simplification, assumed a centrosymmetric electron distribution in calculation by adopting a form of trigonometric or delta function.[22,27,28] However, such an assumption ignores the lopsided shape of electron distribution that originates from the asymmetrically electric gating in single-gated configuration,[26,29] which may result in large calculation inaccuracy. To improve it, we proposed an empirical method to deal with the lopsidedness in electron distribution in atomically thin MoS$_2$;[29] notwithstanding, further researches are still desired to address the accurate electron distribution within semiconductors from the basic physical principles, in particular in the dimensionality crossover regime from 2D to bulk, and, more importantly, to clarify its impacts on macroscopic device performance.

Here, we performed a systematic study on the $\Psi(z)$ variation within the dimensionality-crossed MoS$_2$ channels ranging from 1 to 10 layers by self-consistently solving the 1D Schrödinger-Poisson equations[30] with modulating multiple parameters, including channel thickness ($t$), relative permittivity ($\varepsilon_r$) of channel, electron density ($n_s$) and gate configuration (single- or dual-gated, SG or DG). The calculation shows that, for SG channels, the peak of electron distribution will shift closer to dielectric-channel interface with increasing vertical gating electric field (determined by $n_s$ and $\varepsilon_r$) or decreasing $t$. Before channels entering strong inversion (high $n_s$) or shirking to atomically thickness, we found that $d_p$ decreases as $\varepsilon_r$ decreases or $n_s$ increases, because both parameters can enhance the gating electric field inside channels. The electron distribution under the DG configuration with symmetric gate dielectrics was also investigated, featuring symmetric $\Psi(z)$ forms with one or two peaks (dependent on $t$).

Based on the accurate electron distribution from the Schrödinger-Poisson solutions, the impacts of electron distribution variation on the quantum screening effect among electrons were further studied. Also, we investigated the dependence of individual scattering mechanisms [interfacial Coulomb impurities (CIs), surface optical (SO) phonons, and lattice vacancies] on channel thickness in unit of number of layers (NL) and on overall carrier mobilities at a typical $n_s$ level ~$10^{13}$ cm$^{-2}$ for MoS$_2$ channels encapsulated in hexagonal boron nitride (BN) and HfO$_2$. By comparing the mobility values under SG and DG configurations, we uncovered the possibility in enhancing carrier mobility through engineering the carrier distribution via symmetrizing gate configuration in channels thicker than 3 nm, where the rate of increase reaches up to 23% in HfO$_2$ dual-gated 10-layer channels.

**Methods and Discussion**

To look insight into the effect of gate configuration, we first study $\Psi(z)$ in a 6.5 nm (10 layer) thick $MoS_2$ channel under varied SG and DG configurations, where the channel is encapsulated by two 10 nm thick dielectrics with a $\varepsilon_r$ of 10 and the both dielectrics are covered with metallic gates. Figure 1a shows $\Psi(z)$ of channel electrons at $n_s = 5 \times 10^{12}$ cm$^{-2}$ and corresponding potential profile $V(z)$ of the device under the SG configuration. In such a configuration, the channel and top gate (TG) are electrically grounded while the bottom gate (BG) is biased with a positive voltage $V_{BG}$, as shown in the inset. In comparison with the flat band exhibited by the top dielectric, the band of the gated channel shows a downward bending of 185 meV near the bottom interface, resulting in an electric field of 0.9 MV/cm. Because of the attraction of gate electric field to induced electrons, $\Psi(z)$ is lopsidedly distributed toward the side of BG. To describe such a lopsided distribution, a quantity $d_p$ is introduced to characterize the distance from the distribution peak to the gating dielectric interface. A low $d_p$ value of 1.08 nm is estimated under the SG configuration, indicating the strong interaction and tight coupling of the channel electrons to the dielectric interface.

For comparison, Figure 1b shows $\Psi(z)$ and $V(z)$ under the DG configuration. Here an equivalent-field principle is adopted to compare the electron distributions in channels under the two gate configurations. Hence, a gate voltage of the same magnitude is additionally applied to TG and thus $n_s$ reaches $10^{13}$ cm$^{-2}$, being twice as much as that under the SG configuration. Because the device and biasing conditions become strictly symmetric, $\Psi(z)$ and $V(z)$ are both symmetrized accordingly. The band bending at the channel interfaces is reduced to 158 meV, slightly lower than that in the SG configuration, with its energy maximum located at the center. Likewise, the envelope function becomes centrosymmetric, in spite of the two peaks caused by the electrostatic effect from the dual gates. In this case, $d_p$ is increased to 1.65 nm, indicating the higher possibilities for electrons to occupy the middle region of the channel. As a result, the electrons are more immune to the scattering events from the channel/dielectric interfaces, constituting the rationale for mobility enhancement.

Apart from gate configuration, $t$, $\varepsilon_r$, and $n_s$ are also parameters of interest affecting electron distribution. We then study the dependence of $d_p$ on $n_s$ and $\varepsilon_r$ in channels with different $t$ values of 6.5 (10 layer) and 0.65 nm (monolayer). The results are plotted in Figure 2a. Under the SG configuration, the general tendency for both channels is that $d_p$ decreases with increasing $n_s$ or decreasing $\varepsilon_r$, which is attributed to the reinforcement of the surface electric field given by the equation, $E = en_s/(\varepsilon_0\varepsilon_r)$, where $e$ is the elementary charge and $\varepsilon_0$ is the vacuum dielectric constant. Nevertheless, the rate of variation is different in the two channels, where the dependence of $d_p$ on $n_s$ and $\varepsilon_r$ are more prominent in the thick one. For instance, at $n_s = 10^{13}$ cm$^{-2}$, $d_p$ is 1.14 nm when $\varepsilon_r = 12$ but decreases to 0.7 nm when $\varepsilon_r = 3$. Also, the variation rate of $d_p$ by $n_s$ is quite high, reaching ~60% for $\varepsilon_r = 3$. By contrast, $d_p$ variation is rather small in the thin channel and $d_p$ is almost unchanged as $n_s$ or $\varepsilon_r$ changes (bottom panel, Figure 2a). In the parametric ranges covered, it varies by 3% at most, suggesting that the electron distribution in the monolayer channels is less adjustable because of the spatial confinement effect.

Figure 2b shows the contour plot for $d_p$ versus $t$ (0.65–6.5 nm) and $\varepsilon_r$ (2–13) at a fixed high $n_s$ of $10^{13}$ cm$^{-2}$ under the SG configuration. Three distinct regions ($t$-dependent, $\varepsilon_r$-dependent, and saturated) are seen in the plot, as divided by the dashed lines. In the $t$- dependent region, $d_p$ increases monotonically with increasing $t$ because of the weakening of spatial confinement. Such $t$-dependent behavior occurs in low-$t$ range and persists no more than 4 nm even at high $\varepsilon_r$, while it shifts to $\varepsilon_r$-dependent when $t$ is large enough. In the $\varepsilon_r$-dependent region, $d_p$ is insensitive to $t$ but highly depends on $\varepsilon_r$, varying mainly with the vertical gating electric field. At the saturated region where $\varepsilon_r$ and $t$ are simultaneously high, electron distribution is almost independent on these two parameters and, thus, become almost unadjustable. The above observation is a natural consequence of the Schrödinger-Poisson equations, which include the two coexisting quantum and electrostatic mechanisms that govern the distribution of electrons in channels. In thick channels, the electrons are primarily confined by the band bending caused by the electrostatic gating while, in thin channels, the electrons are squeezed within narrow spaces and the spatial confinement effect becomes dominant.

Technologically speaking, DG FETs can exhibit stronger electrostatic control capacity and feature higher downscaling ability than the SG counterparts.[31] Thus, it is important to contrastively study the electron distributions in the DG channels, in particular in those few-nanometer thick. As shown in Figure 1b, $\Psi(z)$ is centrosymmetric in the DG channels due to symmetrized $V(z)$ and, thus, it would exhibit one or two distribution peaks, depending on $t$. To focus on this feature, here we choose another quantity $d_g$ to describe the gap between the two separated peaks, which is given by $d_g = t - 2d_p$. Note that $d_g$ would be reduced to zero in thin channels or under weak electrostatic modulation where the two peaks evolves into an overlapped one and, thus, $d_p = t/2$ if $d_g = 0$ (i.e., single peak).

Figure 2c presents dependence of $d_g$ on $n_s$ for a 5.2 nm thick (8 layer) DG MoS$_2$ channel with different $\varepsilon_r$ values of 2, 6, and 12. In the low $n_s$ regime where the channel is insufficiently gated, only one overlapped peak appears in the electron distribution and results in $d_g = 0$. However, when $n_s$ increases to some threshold values (dependent on $\varepsilon_r$), two distribution peaks would arise and lead to a non-zero $d_g$ that increases with increasing $n_s$. The threshold values are 1.7, 5.4, and 11.1×10$^{12}$ cm$^{-2}$ for devices with $\varepsilon_r = 2$, 6, and 12, respectively. The corresponding surface electric fields amount to 0.79, 0.82, and 0.84 MV/cm, respectively, suggesting the trend of a high surface electric field required at large $\varepsilon_r$ to divide the electron distribution into two peaks.

To understand the impact of gate configuration on $\Psi(z)$, in Figure 2d we compare $d_p$ between the DG and SG channels with $t = 6.5$ nm (10 layer) and $\varepsilon_r = 10$ in a $n_s$ range of 0–1.5×10$^{13}$ cm$^{-2}$. At the same $n_s$ levels (i.e., $V_{SG} = 2V_{DG}$), $d_p$ in the BG channel is generally smaller than that of DG channel (See the solid black and red lines). Such a trend holds even under the constant-field principle by setting $V_{SG} = V_{DG}$ (i.e., $n_{DG} = 2n_{SG}$), which is equivalent to double scale the x-axis of the solid to the dashed black line. The generally reduced $d_p$ behavior under the DG working configuration is a consequence of the symmetrized electrostatic interaction by gate fields, further confirming the potential immunity for electrons to extrinsic scatterings from interfaces under the DG configuration.

It is well known that quantum screening arises mainly from the electron-electron interaction and behaves as a modulation on response to external electric fields by electron gases in solids.[24,28] In principle, any changes in external stimuli would cause consequent variation in the average distance between electrons and resultant polarization of the dense electron gases confined in the ultrathin channels. Considering that the problem of permittivity mismatch between channels and dielectrics has been well-studied[22,28,29,32] and, for the sake of simplicity, we assume the MoS$_2$ channels are encapsulated in two dielectric layers with the same $\varepsilon_r$ as MoS$_2$. Such dielectric environment without permittivity mismatch greatly simplifies the calculation on scattering matrix elements but preserves the effect of electron distribution variation.

To further shed light on the impacts of electron distribution on carrier scattering, we next study the dependence of quantum screening (i.e., electron-electron interaction) and individual scattering mechanisms on $t$, $n_s$, and gate configuration in nanoscale MoS$_2$ channels. Among the various scattering mechanisms including interfacial CIs,[29] SO phonon,[22,33] surface roughness,[13] and lattice defects,[16] the former two are sensitive to electron distribution. Thus, we focus on the impacts of electron distribution on these two mechanisms.

The dielectric function of electron gas is given by the well-known Lindhard formula[24,34]

$$\varepsilon_{2d}(q, T) = 1 + \frac{e^2}{2\varepsilon_0 \varepsilon_r q} F_{ee}(q) \Pi(q, T), \tag{1}$$

where $q$ is the scattering wave vector of the 2D electrons, $\Pi(q,T)$ is the polarizability function at temperature $T$, and $F_{ee}$ is the form factor given by

$$F_{ee}(q) = \int_0^t \int_0^t \Psi(z_1)^2 \Psi(z_2)^2 \exp(-q|z_1 - z_2|) \, dz_1 dz_2. \tag{2}$$

Figure 3a shows the dependence of $F_{ee}(k_F)$ on $t$ in SG and DG MoS$_2$ at $n_s = 10^{13}$ cm$^{-2}$, where $k_F$ is the Fermi wave vector. According to Equation (2), one can find that $F_{ee}$ is strongly associated with the compactness of electron distribution (i.e., average distance between electrons). In the extremely compact case with an ideal delta distribution $F_{ee} = 1$, while $F_{ee}$ is less than 1 in less compact distributions. In practical cases, the monolayer MoS$_2$ ($t = 0.65$ nm) has the most compact distribution among all thicknesses and, thus, shows the highest $F_{ee}$ value of 0.88. Generally, $F_{ee}$ decreases monotonically with increasing $t$ because of the broadening of electron distribution. However, the rates of decay are different between SG and DG configurations. Under the SG configuration $F_{ee}$ saturate to 0.65 at $t \approx 5$ nm, while it shows no sign of saturation untill 6.5 nm under the DG configuration. Moreover, $F_{ee}$ under the DG configuration is always smaller than that under the SG configuration, because electrons are generally dispersed in the former case. The inset of Figure 3a also shows the dependence of $F_{ee}$ on $q$ for a 6.5 nm channel. In general, $F_{ee}$ decreases monotonically with $q$, and it decreases faster under DG than under SG configuration.

As mentioned, the variation of $n_s$ will also result in the change of average distance between electrons and overall screening effect, we then study the dependence of $F_{ee}$ on $n_s$ at an example $q = 0.5$ nm$^{-1}$ for a 6.5 nm MoS$_2$, as shown in Figure 3b. Interestingly, $F_{ee}$ of the SG channel increases monotonically with increasing $n_s$ in the range of $10^{12}$–$10^{13}$ cm$^{-2}$, while it shows an opposite trend in the DG channel, resulting from the less compact electron distribution in channel under the DG configuration.

For 2D semiconductors, the scattering from interfacial CIs, which originate from dangling bands and chemical residues at channel surfaces, constitutes an important limiting factor for carrier mobility.[29] Hence, we study the effects of variation of electron distribution on its intensity. The scattering matrix element of interfacial CI scattering is given by

$$M_q^{CI} = \frac{e^2}{2S\varepsilon_0 \varepsilon_r q} \frac{F_{ec}(q)}{\varepsilon_{2d}(q)}, \tag{3}$$

where $S$ is the area of channel, $F_{ec}$ is the coupling factor used to evaluate the Coulomb interaction between channel electrons and interfacial CIs. Considering the lopsided distribution of SG channels, $F_{ec}$ contains two terms: $F_{ec}^{BCI}$ and $F_{ec}^{TCI}$ that account for the interaction of electrons to the CIs located at the bottom and top interfaces, respectively.[20,29] For the DG channels, the two terms are the same because of the symmetrized electron distributions.

Likewise, we also calculate the dependence of the two $F_{ec}$ terms on $t$ and $q$, as shown in Figure 3c, which intentionally separates the contributions from the top and bottom interfaces under the SG configuration. All these terms decrease monotonically with $t$, because the interaction distances (scaling roughly with $d_p$ and $t-d_p$) between electrons and interfacial CIs increase as increasing $t$. Still assuming the SG channel is bottom gated, we find that $F_{ec}^{BCI}$ saturates to ~0.47 around $t \approx 5$ nm due to the saturation of $d_p$. By contrast, $F_{ec}^{TCI}$ keeps decreasing till $t = 6.5$ nm, because the distance between electrons and top interfacial CIs (i.e., $t-d_p$) will not saturate with increasing $t$, which renders the scattering from the top interfacial CIs negligible in thick channels. $F_{ec}$ under DG configuration is smaller than $F_{ec}^{BCI}$ but larger than $F_{ec}^{TCI}$ under SG counterpart because $d_p$ in the former case is always in the middle of the latter two cases.

The inset in Figure 3c shows the $q$ dependence of the three $F_{ec}$ terms for a 6.5 thick MoS$_2$ at $n_s = 10^{13}$ cm$^{-2}$. All the terms sharply decay with increasing $q$, roughly following an exponential law, $F_{ec} = \exp(-q d_{eq})$, where $d_{eq}$ is an equivalent distance between interfacial CIs and channel electrons that are projected to obey an ideal delta distribution. For the SG channel with $n_s = 10^{13}$ cm$^{-2}$, the $d_{eq}$ for the bottom (top) interface is about 0.53 (2.0) nm. For the DG channel, the $d_{eq}$ is estimated to be 0.96 nm. This result can be used as a fast approximation to calculate the complicated $F_{ec}$ terms.[20,29] Also, Figure 3d shows dependence of $F_{ec}$ terms on $n_s$ at the example $q$ of 0.5 nm$^{-1}$. Under the DG configuration $F_{ec}$ is weakly dependent on $n_s$, which only increases from 0.24 to 0.26 in the broad $n_s$ range from $10^{12}$ to $10^{13}$ cm$^{-2}$. By contrast, the increasing or decreasing trend of the $F_{ec}^{BCI}$ or $F_{ec}^{TCI}$ term is more appreciable under the SG configuration, resulting from the dramatic variation of the lopsided electron distribution in the SG channels.

After understanding the impacts of device parameters ($t$ and $n_s$) and gate configuration on crucial terms $\Psi(z)$, $F_{ee}$ and $F_{ec}$, we further look into insight on their influence on the CI-limited mobility ($\mu_{CI}$). By assuming a same CI density of $10^{12}$ cm$^{-2}$ at the top and bottom interfaces in a 6.5-nm MoS$_2$ channel,[20] we calculated $\mu_{CI}$ under the two gate configurations on the basis of the accurate electron distributions from the Schrödinger-Poisson solutions, as shown by the red and blue lines in Figure 4a. Under both cases, $\mu_{CI}$ decreases with increasing $n_s$. Such a descending trend is mainly attributed to the negative correlation between $d_p$ and $n_s$ (Figure 2d), which can increase the Coulomb interaction between channel electrons and interfacial CIs. Besides, the values of $\mu_{CI}$ under the SG configuration are much smaller than those under the DG configuration, because of the generally smaller $d_p$ at a fixed $n_s$ under the former case.

For comparison with the conventional trigonometric approximation on $\Psi(z)$, we also calculated the relevant $\mu_{CI}$–$n_s$ curve (black line in Figure 4a), which exhibits a slight ascending trend with $n_s$. Such an ascending trend is an overall consequence due to the enhancement of quantum screening and the increase of Fermi level as $n_s$ increases, but it contradicts with the calculations with realistic electron distributions, indicating the presence of large deviation in calculating electronic performance of thick channels by using the conventional trigonometric $\Psi(z)$.

In Figure 4b, we further clarify the inconsistency induced by inaccurate $\Psi(z)$ through fully calculating the $\mu_{CI}$–$t$ curves for the accurate and trigonometric electron distributions. In all cases, the $\mu_{CI}$–$t$ trends are consistent. However, the $\mu_{CI}$ curves based on the accurate distributions saturate more easily in thick channels due to the fast saturation of electron distribution, while the curve based on the trigonometric distribution shows no sign to saturate. Besides, the $\mu_{CI}$ would be generally overestimated in channels thicker than 3 nm and the overestimation reaches up to an intolerable level of 3 folds at $t > 5$ nm when calculated from the approximation of trigonometric distribution. Hence, the trigonometric approximation is acceptable only for channels below 3 nm, representing useful guidance for device modeling.

It has been reported that the scattering from SO phonons in high-k dielectrics represents another significant interfacial scattering mechanism relevant to carrier mobility.[21,33] The scattering matrix element of SO phonon scattering can be written as[33,35]

$$M_q^{SO} = \frac{F_{SO}(q)}{\varepsilon_{2d}(q)} \sqrt{\frac{e^2 E_{SO}}{2Sq\varepsilon_0} \left( \frac{1}{\varepsilon_{ox}^\infty + \varepsilon_r^\infty} - \frac{1}{\varepsilon_{ox}^0 + \varepsilon_r^\infty} \right)}, \tag{4}$$

where $E_{SO}$ is the energy of SO phonon, $\varepsilon_{ox}^\infty$ ($\varepsilon_{ox}^0$) is the optical (static) permittivity of the dielectric, and $F_{SO}$ is the coupling factor to weight the interaction between electrons and SO phonons. Likewise, we use $F_{SO}^{BCI}$ and $F_{SO}^{TCI}$ to denote the SO phonon scattering from the bottom and top dielectrics, respectively.[22,24,36]

In addition to interfacial CIs and SO phonons, structural defects (DFs) in MoS$_2$ lattice could be a third important type of leading scattering mechanism.[23,37] Although there are many categories of structural defects in MoS$_2$, the lattice vacancies due to the loss of sulfur atoms at channel surfaces are believed to be the dominant one to be considered in high-quality samples.[38,39] Normally, the lattice vacancies are treated as short-range scattering mechanism[16,29] and, thus, we assume the lattice vacancies are mainly distributed in the surfacial layers of MoS$_2$ by adopting a delta approximation for the scattering potential as $eL_B^2/(\varepsilon_0\varepsilon_r)\,\delta(x, y, z - z_0)$, where $L_B$ is the length of Mo-S bond. As such, the scattering matrix element is derived as

$$M_q^{DF} = \frac{e^2 L_B^2 \Psi(z_0)^2}{\varepsilon_{2d}(q)\varepsilon_0\varepsilon_r S}. \tag{5}$$

Finally, we calculate the dependence of room-temperature mobility on channel thickness in unit of NL under different gate configurations and with varied dielectrics based on the accurate electron distributions. For accuracy, the mismatched dielectric environments are strictly considered in our calculation.[22,40] Figure 5 shows the mobilities of BN- and HfO$_2$-encapsulated MoS$_2$ under the SG and DG configurations as a function of NL. The leading scattering sources are also plotted individually, including intrinsic phonons, CI, SO phonons, and DFs. In case of BN dielectric (Figure 5a), the component of SO phonons is neglected because it is too weak to be considered.[22,41] The CI densities ($n_{CI}$) at the bottom and top interfaces are assumed to be identical with $n_{CI} = 2 \times 10^{11}$ cm$^{-2}$ for BN [42] and $10^{12}$ cm$^{-2}$ for HfO$_2$, respectively. The densities of defects ($n_{DF}$) are assumed to be $10^{13}$ cm$^{-2}$ at both channel surfaces.[37,38] The intrinsic mobility ($\mu_{int}$) from lattice phonons is adopted from pervious prediction,[43] having a NL-independent value of 410 cm$^2$V$^{-1}$s$^{-1}$ at room temperature. The total mobilities are estimated with the Matthiessen's rule.

Two important pieces of information for performance optimization can be seen from Figure 5. First, $\mu$ increases with NL because the CI, DF, and SO limited mobilities all show positive correlation with NL, resulting from the general large interaction distance $d_p$ between electrons and surface scattering sources. From the point of view of mobility only, the channels with few-layer thick are more appropriate than monolayers for more-Moore electronics, despite the extraordinary immunity of monolayers to the short-channel effect. For applications as driving transistors where the downscaling is not so urgent, slightly thick DG channels (> 5 layers) are more favorable.

Second, $\mu$ components are generally higher under the DG than SG configuration for all the three surface associated mechanisms, which is also associated with the variation in compactness of electron distribution due to symmetrized electrostatic gating under the DG configuration. Although the difference in mobility is inappreciable between the DG and SG configuration when NL < 5, it quickly increases in thicker channels with NL > 5. In case of $HfO_2$ encapsulated FETs, gate configuration can bring about 23% enhancement in overall mobility at NL = 10. Therefore, we propose that the carrier mobility in nanoscale semiconductors can be effectively enhanced through appropriately engineering the carrier distributions in channels via optimizing thickness and gate configuration.

It is noteworthy that the above results are calculated for high-quality devices because all the parameters of the extrinsic scattering sources are adopted at low levels, where the impacts of electron distribution variation are minimized because the electron distribution irrelevant lattice phonons act as the primary mobility limit in the regime of thick channels. For most practical devices, the levels of CIs and DFs can be even 1–2 orders higher in magnitude and they likely serve as the dominant scattering sources. The enhancement rates of mobility through the carrier distribution engineering can be higher under such circumstances.

In summary, we have provided a deep insight into the relationship of electron distributions with multiple factors including gate configuration, channel thickness, dielectric permittivity, and electron density, by virtue of self-consistent solutions from the 1D Schrödinger-Poisson equations. Also, we have clarified its impacts on various carrier scattering matrix elements and overall carrier mobility based on Boltzmann transport equation. Importantly, we show the possibility in enhancing carrier mobility in 2D semiconductors through engineering the vertical distribution of carriers confined inside the ultrathin channels via symmetrizing gate configuration or increasing channel thickness. The rate of enhancement reaches up to 23% in $HfO_2$ dual-gated clean $MoS_2$ channels. These systematic results represent useful guidance for device design and performance optimization in developing ultrathin transistors for advanced electronics.

### Acknowledgements

This work was supported by the National Key R&D Program of China (2021YFA1202903), the National Natural Science Foundation of China (92264202, 61974060 and 61674080), the Innovation and Entrepreneurship Program of Jiangsu province.

### References

[1] Shen P C, Su C, Lin Y, Chou A S, Cheng C C, Park J H, Chiu M H, Lu A Y, Tang H L, Tavakoli M M, Pitner G, Ji X, Cai Z, Mao N, Wang J, Tung V, Li J, Bokor J, Zettl A,

Wu C I, Palacios T, Li L J and Kong J 2021 *Nature* **593** 211–217

[2] Desai S B, Madhvapathy S R, Sachid A B, Llinas J P, Wang Q, Ahn G H, Pitner G, Kim M J, Bokor J, Hu C, Wong H S P and Javey A 2016 *Science* **354** 99–102

[3] Akinwande D, Huyghebaert C, Wang C H, Serna M I, Goossens S, Li L J, Wong H S P and Koppens F H 2019 *Nature* **573** 507–518

[4] Quhe R, Xu L, Liu S, Yang C, Wang Y, Li H, Yang J, Li Q, Shi B, Li Y, Pan Y, Sun X, Li J, Weng M, Zhang H, Guo Y, Xu L, Tang H, Dong J, Yang J, Zhang Z, Lei M, Pan F and Lu J 2021 *Phys. Rep.* **938** 1–72

[5] Meng W, Xu F, Yu Z, Tao T, Shao L, Liu L, Li T, Wen K, Wang J, He L, Sun L, Li W, Ning H, Dai N, Qin F, Tu X, Pan D, He S, Li D, Zheng Y, Lu Y, Liu B, Zhang R, Shi Y and Wang X 2021 *Nat. Nanotechnol.* **16** 1231–1236

[6] Shin J, Kim H, Sundaram S, Jeong J, Park B I, Chang C S, Choi J, Kim T, Saravanapavanantham M, Lu K, Kim S, Suh J M, Kim K S, Song M K, Liu Y, Qiao K, Kim J H, Kim Y, Kang J H, Kim J, Lee D, Lee J, Kim J S, Lee H E, Yeon H, Kum H S, Bae S H, Bulovic V, Yu K J, Lee K, Chung K, Hong Y J, Ougazzaden A and Kim J 2023 *Nature* **614** 81–87

[7] Hwangbo S, Hu L, Hoang A T, Choi J Y and Ahn J H 2022 *Nat. Nanotechnol.* **17** 500–+

[8] Ieong M, Doris B, Kedzierski J, Rim K and Yang M 2004 *Science* **306** 2057–2060

[9] Yan R H, Ourmazd A and Lee K 1992 *IEEE Trans. Electron Dev.* **39** 1704–1710

[10] Hisamoto D, Kaga T, Kawamoto Y and Takeda E 1989 A fully depleted lean-channel transistor (DELTA)-a novel vertical ultrathin SOI MOSFET *IEEE International Electron Devices Meeting* (Washington, DC, USA: IEEE) pp 833–836

[11] He X, Fronheiser J, Zhao P, Hu Z, Uppal S, Wu X, Hu Y, Sporer R, Qin L, Krishnan R, Bazizi E M, Carter R, Tabakman K, Jha A K, Yu H, Hu O, Choi D, Lee J G, Samavedam S B and Sohn D K 2017 Impact of aggressive fin width scaling on FinFET device characteristics *IEEE International Electron Devices Meeting* (San Francisco, CA: IEEE) pp 2021–2024

[12] Bhoir M S, Chiarella T, Ragnarsson L Å, Mitard J, Terzeiva V, Horiguchi N and Mohapatra N R 2019 *IEEE J. Electron Devices Soc.* **7** 1217–1224

[13] Jin S, Fischetti M and Tang T W 2007 *IEEE Trans. Electron Dev.* **54** 2191–2203

[14] Gomez L, Aberg I and Hoyt J L 2007 *IEEE Electr. Device Lett.* **28** 285–287

[15] Fiori G, Bonaccorso F, Iannaccone G, Palacios T, Neumaier D, Seabaugh A, Banerjee S K and Colombo L 2014 *Nat. Nanotechnol.* **9** 768–779

[16] Cui X, Lee G H, Kim Y D, Arefe G, Huang P Y, Lee C H, Chenet D A, Zhang X, Wang L, Ye F, Pizzocchero F, Jessen B S, Watanabe K, Taniguchi T, Muller D A, Low T, Kim P and Hone J 2015 *Nat. Nanotechnol.* **10** 534–540

[17] Radisavljevic B and Kis A 2013 *Nat. Mater.* **12** 815–820

[18] Chhowalla M, Jena D and Zhang H 2016 *Nat. Rev. Mater.* **1** 16052

[19] Liu Y, Duan X, Shin H J, Park S, Huang Y and Duan X 2021 *Nature* **591** 43–53

[20] Ju S, Liang B, Zhou J, Pan D, Shi Y and Li S 2022 *Nano Lett.* **22** 6671–6677

[21] Zeng L, Xin Z, Chen S, Du G, Kang J and Liu X 2013 *Appl. Phys. Lett.* **103** 113505

[22] Ma N and Jena D 2014 *Phys. Rev. X* **4** 011043

[23] Li S L, Tsukagoshi K, Orgiu E and Samori P 2016 *Chem. Soc. Rev.* **45** 118–151

[24] Ando T, Fowler A B and Stern F 1982 *Rev. Mod. Phys.* **54** 437–672

[25] Fang F and Howard W 1966 *Phys. Rev. Lett.* **16** 797

[26] Kim S, Konar A, Hwang W S, Lee J H, Lee J, Yang J, Jung C, Kim H, Yoo J B, Choi J Y, Jin Y W, Lee S Y, Jena D, Choi W and Kim K 2012 *Nat. Commun.* **3** 1011

[27] Jena D and Konar A 2007 *Phys. Rev. Lett.* **98** 136805

[28] Ong Z Y and Fischetti M V 2013 *Phys. Rev. B* **88** 165316

[29] Li S L, Wakabayashi K, Xu Y, Nakaharai S, Komatsu K, Li W W, Lin Y F, Aparecido-Ferreira A and Tsukagoshi K 2013 *Nano Lett.* **13** 3546–3552

[30] Stern F 1970 *J. Comput. Phys.* **6** 56–67

[31] Colinge J P (ed) 2008 *FinFETs and other Multi-Gate Transistors* (Boston, MA: Springer) ISBN 978-0-387-71751-7

[32] Trolle M L, Pedersen T G and Véniard V 2017 *Sci. Rep.* **7** 39844

[33] Fischetti M V, Neumayer D A and Cartier E A 2001 *J. Appl. Phys.* **90** 4587–4608

[34] Maldague P F 1978 *Surf. Sci.* **73** 296–302

[35] Wang S Q and Mahan G D 1972 *Phys. Rev. B* **6** 4517–4524

[36] Konar A, Fang T and Jena D 2010 *Phys. Rev. B* **82** 115452

[37] Yu Z, Pan Y, Shen Y, Wang Z, Ong Z Y, Xu T, Xin R, Pan L, Wang B, Sun L, Wang J, Zhang G, Zhang Y W, Shi Y and Wang X 2014 *Nat. Commun.* **5** 5290

[38] Hong J, Hu Z, Probert M, Li K, Lv D, Yang X, Gu L, Mao N, Feng Q, Xie L, Zhang J, Wu D, Zhang Z, Jin C, Ji W, Zhang X, Yuan J and Zhang Z 2015 *Nat. Commun.* **6** 6293

[39] Zhou W, Zou X, Najmaei S, Liu Z, Shi Y, Kong J, Lou J, Ajayan P M, Yakobson B I and Idrobo J C 2013 *Nano Lett.* **13** 2615–2622

[40] Kumar A and Ahluwalia P 2012 *Physica B* **407** 4627–4634

[41] Perebeinos V and Avouris P 2010 *Phys. Rev. B* **81** 195442

[42] Liang B, Wang A, Zhou J, Ju S, Chen J, Watanabe K, Taniguchi T, Shi Y and Li S 2022 *ACS Appl. Mater. Interfaces* **14** 18697–18703

[43] Kaasbjerg K, Thygesen K S and Jacobsen K W 2012 *Phys. Rev. B* **85** 115317

# Figures and captions

## Figure 1

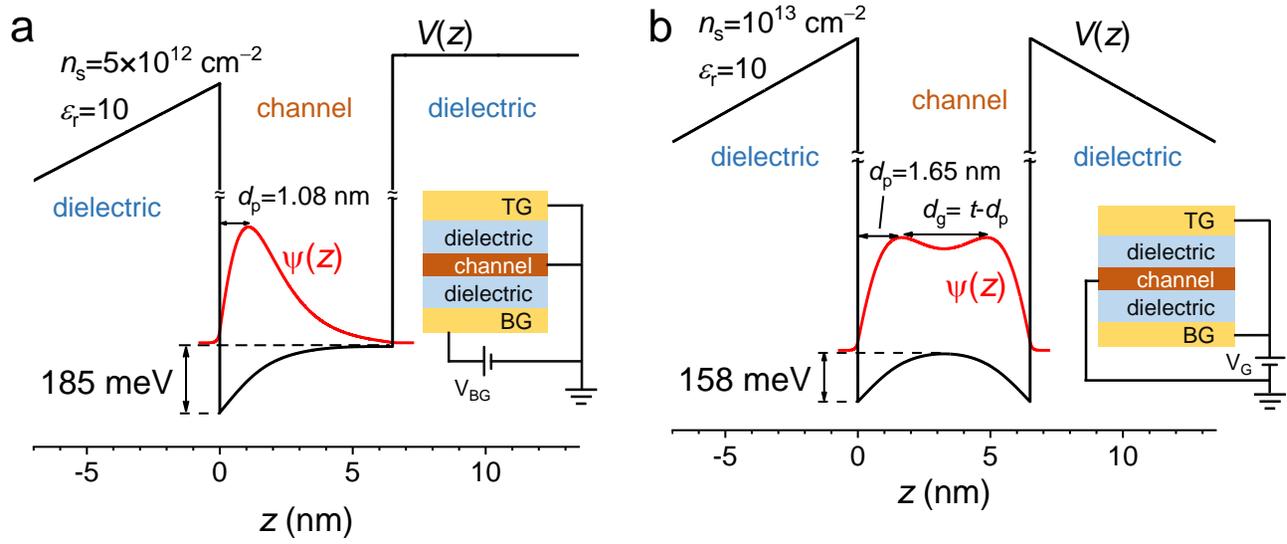

**Figure 1.** Band diagram $V(z)$ and electron envelope function $\Psi(z)$ for (a) a bottom-gated channel with $n_s = 5 \times 10^{12}$ cm$^{-2}$ and (b) a dual-gated channel with $n_s = 10^{13}$ cm$^{-2}$. For the two channels, $t = 6.5$ nm and $\varepsilon_r = 10$. Insets show the corresponding gate configurations.

# Figure 2

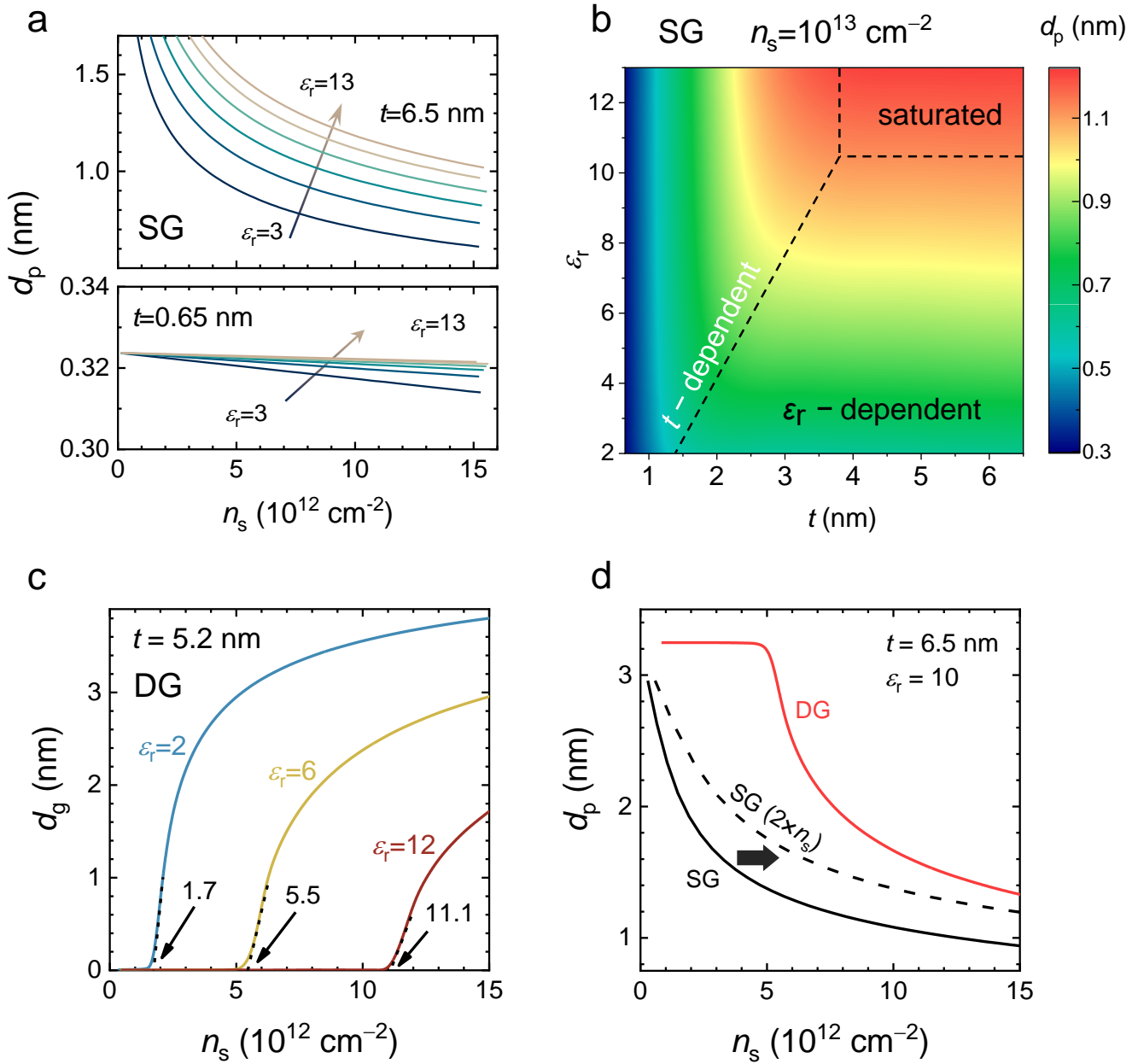

**Figure 2.** (a) Comparison of the dependence of $d_p$ on $n_s$ between a 6.5 and 0.65 nm SG channel with various $\varepsilon_r$ in the ranging from 3 to 13 at a step of 2. (b) Contour plot for $d_p$ versus $t$ and $\varepsilon_r$ at $n_s = 10^{13}$ cm$^{-2}$. (c) Comparison of the dependence of $d_g$ on $n_s$ in a 5.2 nm thick DG channel with $\varepsilon_r = 2$, 6, 12. (d) Comparison of the dependence of $d_p$ on $n_s$ in channels under varied gate configurations. Solid red and black lines denote the cases of DG and SG configurations, respectively. Dashed black line represents the result for a SG channel under the equivalent-field principle (i.e., $V_{SG} = V_{BG}$), which leads to a doubled $n_s$.

# Figure 3

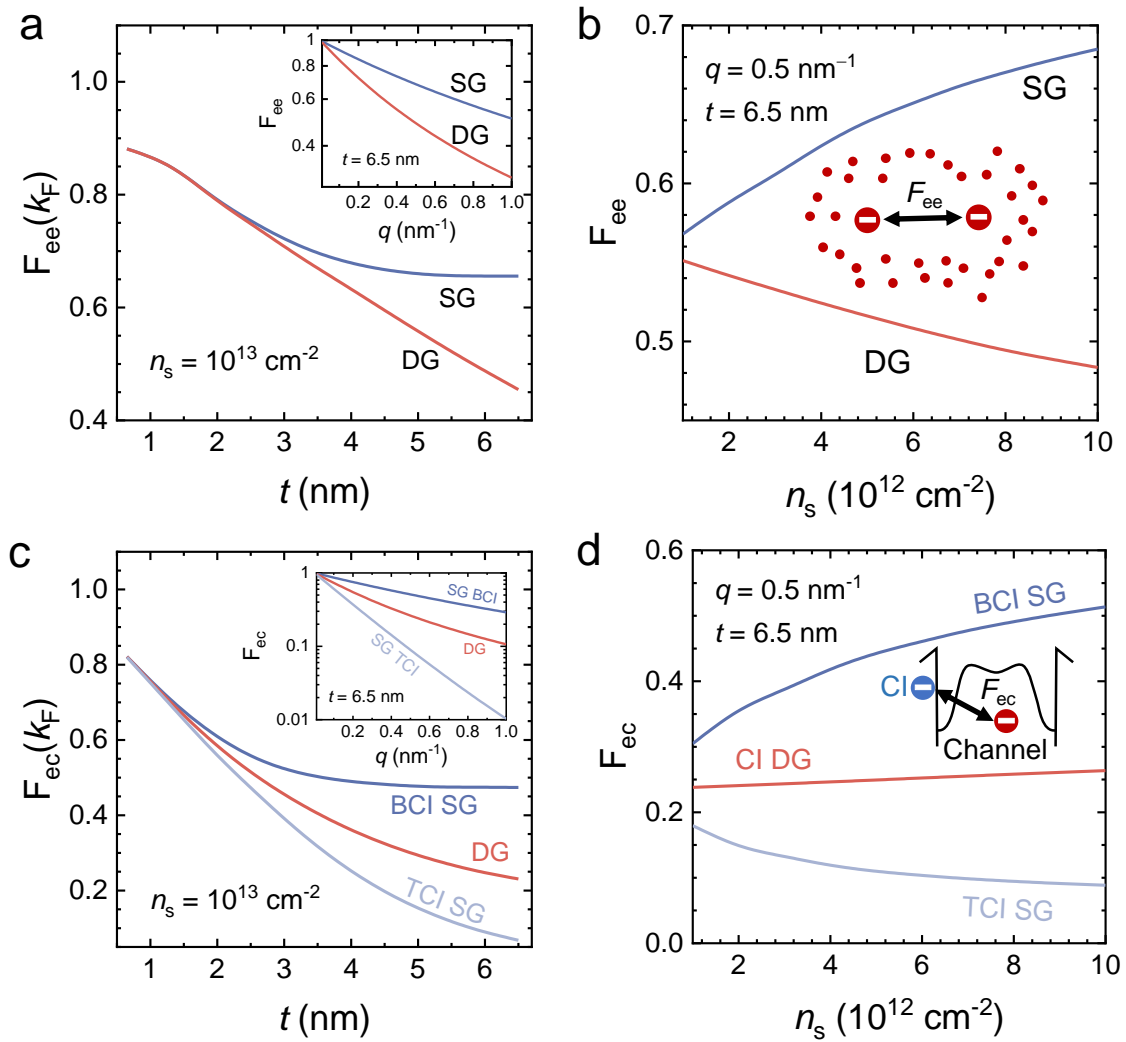

**Figure 3.** Dependence of quantum screening term $F_{ee}$ on (a) $t$ at $q = k_F$ and $n_s = 10^{13}$ cm$^{-2}$, (inset of a) $q$ at t=6.5 nm and $n_s = 10^{13}$ cm$^{-2}$, and (b) $n_s$ at $q = 0.5$ nm$^{-1}$ and $t = 6.5$ nm. Inset of (b) shows the schematic diagram for screened Coulomb interaction between two channel electrons confined in the background of electron sea in the channel. The impacts from gate configuration are also shown. Inset shows the dependence of $F_{ee}$ on $q$ for DG and SG MoS$_2$ with $t = 6.5$ nm. (b) The F$_{ee}$ of SG and DG MoS$_2$ at $q = 0.5$ nm$^{-1}$ as a function $n_s$. (c) and (d) Corresponding plots for the $F_{ec}$ terms due to Coulomb coupling between channel electrons and CIs. The device parameters are adopted same to (a) and (b). Inset of (d) shows the schematic origin for the $F_{ec}$ terms.

# Figure 4

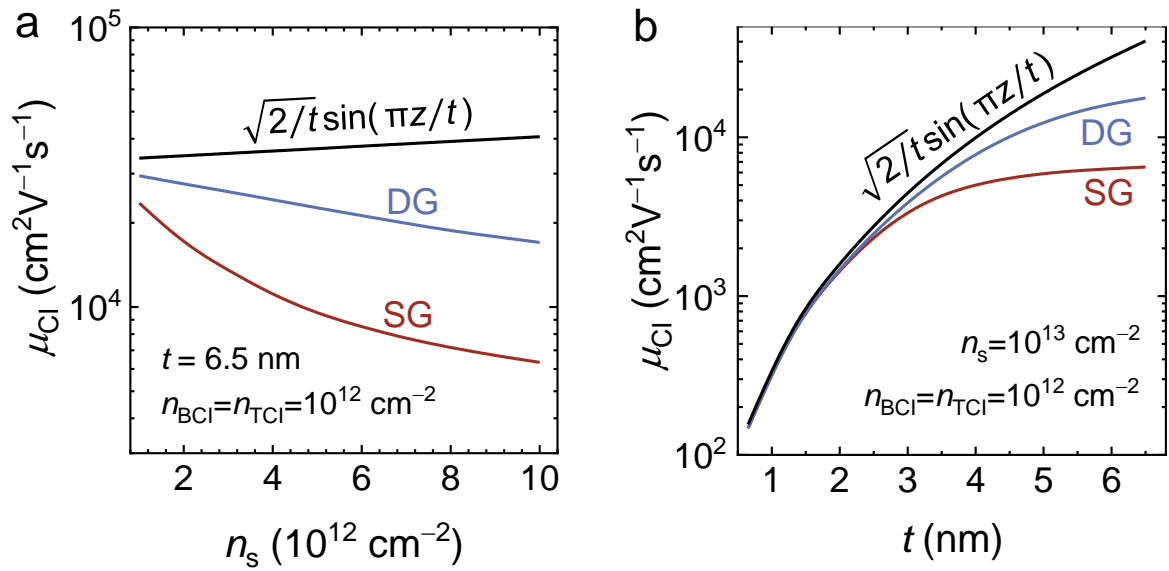

**Figure 4.** (a) Calculated dependence of $\mu_{CI}$ on $n_s$ for SG and DG MoS$_2$ channels with accurate and conventional trigonometric electron distributions. Channel thickness is set to 6.5 nm. (b) Relevant dependence of $\mu_{CI}$ on $t$ at $n_s = 10^{13}$ cm$^{-2}$.

# Figure 5

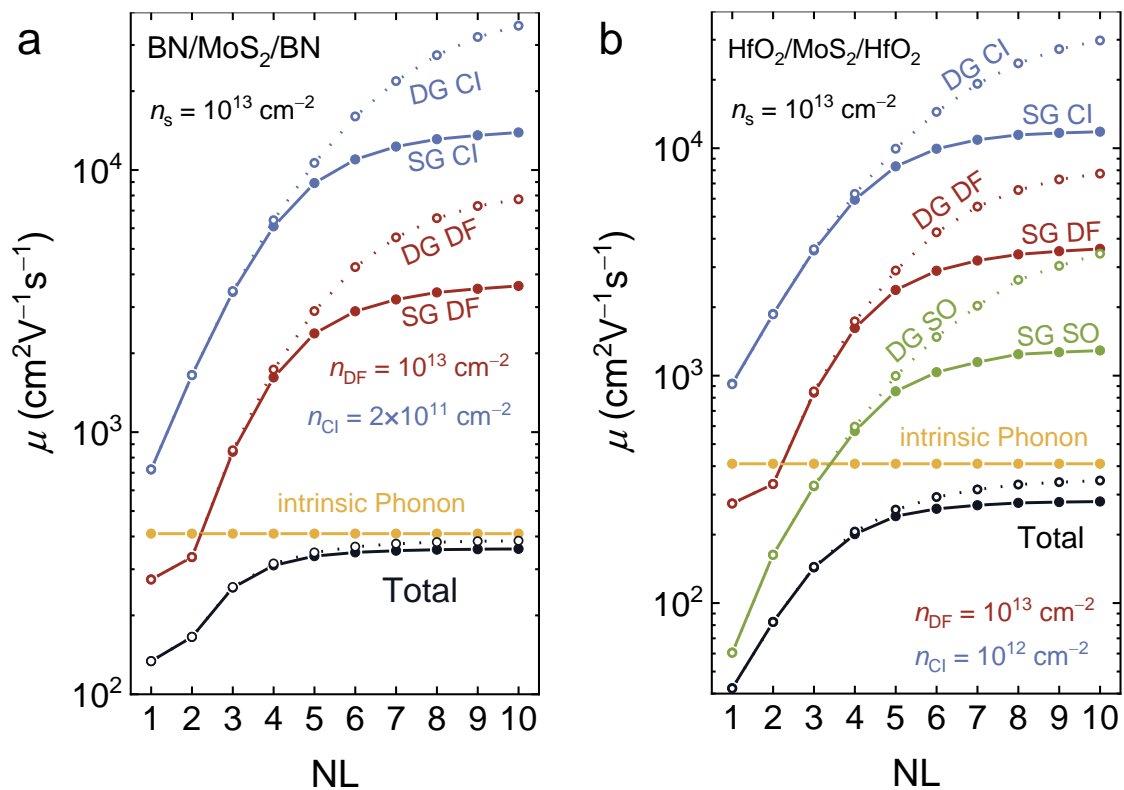

**Figure 5.** Calculated individual $\mu$ components and overall values as a function of NL at room temperature with $n_s = 10^{13}$ cm$^{-2}$ for various scattering mechanisms including CIs, DFs, lattice phonons, and SO phonons in (a) BN-encapsulated MoS$_2$ and (b) HfO$_2$-encapsulated MoS$_2$.